\documentclass[%
reprint,
 amsmath,amssymb,
 aps,
superscriptaddress
]{revtex4-2}

\usepackage{natbib}

\usepackage{graphicx}
\usepackage{dcolumn}
\usepackage{bm}

\usepackage{epsf}
\usepackage{amsmath,amssymb}
\usepackage{slashed}

\usepackage{graphicx}
\usepackage{subfig}

\usepackage{comment}

\usepackage{url}

\usepackage{tikz}
\usetikzlibrary{quantikz2}

\begin{document}

\title{
Quantum Enhancement in Dark Matter Detection with Quantum Computation
}

\author{Shion Chen}
\affiliation{International Center for Elementary Particle Physics (ICEPP), The University of Tokyo, Tokyo 113-0033, Japan}

\author{Hajime Fukuda}
\affiliation{Department of Physics, The University of Tokyo, Tokyo 113-0033, Japan}

\author{Toshiaki Inada}
\affiliation{International Center for Elementary Particle Physics (ICEPP), The University of Tokyo, Tokyo 113-0033, Japan}

\author{Takeo Moroi}
\affiliation{Department of Physics, The University of Tokyo, Tokyo 113-0033, Japan}
\affiliation{QUP (WPI), KEK, Tsukuba, Ibaraki 305-0801, Japan}

\author{Tatsumi Nitta}
\affiliation{International Center for Elementary Particle Physics (ICEPP), The University of Tokyo, Tokyo 113-0033, Japan}

\author{Thanaporn Sichanugrist}
\email[Corresponding author.]{}
\affiliation{Department of Physics, The University of Tokyo, Tokyo 113-0033, Japan}

\begin{abstract}

We propose a novel method to significantly enhance the signal rate in qubit-based dark matter detection experiments with the help of quantum interference. Various quantum sensors possess ideal properties for detecting wave-like dark matter, and qubits, commonly employed in quantum computers, are excellent candidates for dark matter detectors. We demonstrate that, by designing an appropriate quantum circuit to manipulate the qubits, the signal rate scales proportionally to $n_{\rm q}^2$, with $n_{\rm q}$ being the number of sensor qubits, rather than linearly with $n_{\rm q}$. Consequently, in the dark matter detection with a substantial number of sensor qubits, a significant increase in the signal rate can be expected. We provide a specific example of a quantum circuit that achieves this enhancement by coherently combining the phase evolution in each individual qubit due to its interaction with dark matter. We also demonstrate that the circuit is fault tolerant to de-phasing noises, a critical quantum noise source in quantum computers. The enhancement mechanism proposed here is applicable to various modalities for quantum computers, provided that the quantum operations relevant to enhancing the dark matter signal can be applied to these devices.

\end{abstract}  

\maketitle

\vspace{1mm}
\noindent
\section{Introduction}
Quantum devices, which rely on fundamental quantum properties, play pivotal roles in modern physics. The advancements and applications of these quantum devices remain essential for further progress in our field. Presently, one major motivation for the development of quantum devices is to realize quantum computers, where quantum bits (qubits) are utilized for quantum computations. Additionally, the remarkable sensitivity to external fields, particularly, electric and magnetic fields of certain quantum devices including qubits, makes them invaluable as high-precision quantum sensors.

Such quantum sensors can be utilized even for the detection of fields which have never been observed before (for review, see Ref.\,\cite{Chou:2023hcc}). Particularly, recent studies have pointed out the potential use of quantum sensors, such as the superconducting qubit, trapped electrons, and nitrogen vacancy centers, in detecting a specific type of dark matter (DM), referred to as ``wave-like DM'' \cite{Chen:2022quj,Fan:2022uwu,Chigusa:2023hms,Engelhardt:2023qjf} (see also Refs.\,\cite{Kennedy:2020bac,Dixit:2020ymh,Agrawal:2023umy}). Although various cosmological and astrophysical observations strongly suggest the presence of DM in our universe, its particle physics properties remain largely unknown.  Detection of DM is an essential endeavor for comprehending the history of the universe and for advancing our understanding of physics beyond the standard model.

In a recent publication \cite{Chen:2022quj}, we have argued that the transmon qubit \cite{PhysRevA.76.042319} has ideal properties for the detection of a certain type of wave-like DM, such as hidden-photon DM. The wave-like DM may induce the excitation of the qubit, providing a distinct signal of the DM. It has been shown that the search for the hidden photon DM with the transmon qubit can probe the parameter regions that have remained unexplored. Quantum sensors introduce innovative avenues for the DM detection. 

To further improve the detection sensitivity, experiments with a larger number of qubits offer a clear advantage. With multiple qubits, one can independently observe the excitation of each qubit. In such a procedure, the signal rate (i.e., the probability of observing at least one qubit excitation) is roughly proportional to the number of qubits (denoted as $n_{\rm q}$), assuming that the excitation probability of each qubit is much smaller than $n_{\rm q}^{-1}$. Such a scaling is however drastically changed with the quantum-enhanced parameter estimation technique\,\cite{QuantumEnhancedMeasurement, NaturePhoton5-222, PRL115-170801}. With a proper set of quantum operations, the signal rate can become proportional to $n_{\rm q}^2$, which provides a significant improvement of the sensitivity in the DM detection.

The purpose of this Letter is to demonstrate the protocol to achieve such quantum enhancement. The key idea is to coherently sum up the phase evolution inscribed by the DM in each qubit, using a quantum circuit with a set of single- or two-bit gate operations. 
The entanglement between a large number of qubits plays a vital role.
We provide an explicit example of quantum circuits and elucidate the core concept behind the enhancement of the DM signal.
Notably, platforms hosting a large number of qubits and such gate operations are already realized in the cutting-edge noisy intermediate-scale quantum computer machines based on superconducting transmon qubits~\cite{IBM,arute2019quantum} or trapped ions~\cite{ionQ}, which are even on the way of scaling up in terms of both quantity and quality.
This implies that the future or even the present quantum computer machine may be ideal hardware for the  DM detection experiment of our proposal.

Throughout the Letter, we assume a system consisting of an array of superconducting transmon qubits. However, the enhancement protocol should apply also to the other qubit modalities (e.g., ion trap, Rydberg atoms, etc.) if they can be embedded into quantum circuits.

\vspace{1mm}
\noindent
\section{Qubit excitation by DM}
We start with
discussing the evolution of a single qubit under the influence of DM.  We denote the ground and excited states of the qubit as
$\ket{g}$ and $\ket{e}$, respectively, and the energy difference
between the ground and the excited states as $\omega$. We consider the
case that the DM field is oscillating with the frequency equal to its
mass $m_X$ and that such an oscillating DM field results in the following
effective Hamiltonian of the qubit:
\begin{align}
\label{eq:hamiltonian}
  \hat{\cal H} = 
  \omega \ket{e} \bra{e}
  - 2 \eta \cos (m_X t - \alpha)
  ( \ket{e}\bra{g} + \ket{g}\bra{e} ).
\end{align}
Here, $\eta$ is a constant and $\alpha$ is a phase parameter in
association with the DM oscillation.  As discussed in
Ref.\ \cite{Chen:2022quj} in detail, for the DM search with a transmon
qubit, an important example is the hidden photon DM.  The hidden
photon DM induces an effective ac electric field via its kinetic mixing
with the ordinary photon, which results in the effective Hamiltonian
given in Eq.\ \eqref{eq:hamiltonian}.  Denoting the kinetic mixing
parameter as $\epsilon$, the $\eta$ parameter for the case of the
hidden photon DM is given by
\begin{align}
  \eta =
  \frac{1}{2} \epsilon \kappa d \sqrt{C \omega \rho_{\rm DM}}
  \cos\Theta.
\label{eq:eta}
\end{align}
Here, $d$ and $C$ are parameters characterizing the transmon qubit; $d$
is the effective distance between two conductor plates while $C$ is
the capacitance of the conductor.  In addition, $\kappa$ is the
``package coefficient,'' parameterizing the effect of the cavity
surrounding the qubit, $\rho_{\rm DM}$ is the energy density of the
DM, and $\Theta$ is the angle between the hidden photon polarization
and the normal axis of the capacitor plate of the transmom qubit. The value of $\Theta$ is expected to be approximately constant within the coherence timescale of DM, and an average $\cos^2\Theta \rightarrow 1/3$ should be taken for the numerical estimation due to its randomness.

The DM oscillation can induce the excitation process $\ket{g}\rightarrow\ket{e}$.
Thus, the DM search can be performed by observing such an excitation.  For the DM search, the case of $\omega=m_X$, i.e., the
resonant limit, is of particular importance because the excitation
rate is resonantly enhanced in such a limit.  In the following, we
concentrate on the resonant limit.  Denote the single qubit as
\begin{align}
  \ket{\psi (t)} = 
  \psi_g (t) \ket{g}
  + e^{-i\omega t} \psi_e (t) \ket{e},
  \label{eq:interpic}
\end{align}
which evolves as $i \frac{d}{dt} \ket{\psi
(t)} = \hat{\cal H} \ket{\psi (t)}$. For the time interval of $0\leq t\leq\tau$, where $\tau$ is the
coherence time of the system and is the longest time interval during
which the coherence of the system is maintained, the state
experiences the unitary evolution as
\begin{align}
  \left(
  \begin{array}{c}
    \psi_g (\tau) \\ \psi_e (\tau)
  \end{array}
  \right) = 
  U_{\rm DM}
  \left(
  \begin{array}{c}
    \psi_g (0) \\ \psi_e (0)
  \end{array}
  \right).
\end{align}
With the rotation wave approximation, we can obtain
\begin{align}
  U_{\rm DM} \simeq
  \left(
  \begin{array}{cc}
    \cos\delta & ie^{-i\alpha} \sin\delta
    \\ i e^{i\alpha} \sin\delta & \cos\delta
  \end{array}
  \right),
\end{align}
with
\begin{align}
  \delta \equiv \eta \tau.
\end{align}
The eigenvalues of the unitary operator
$U_{\rm DM}$ are given by $e^{\pm i\delta}$, for which
the eigenstates are $\ket{\psi_\pm} \equiv \frac{1}{\sqrt{2}}(\ket{g}
\pm e^{i\alpha}\ket{e})$. Note that, although we assume
this particular form of $U_{\rm DM}$ in this Letter, our result can be
easily generalized to arbitrary interaction between a qubit and the DM.
For convenience, we simply included evolution factor $e^{-i\omega t}$
in the definition of excited state $\ket{e}$, which also apply hereafter.

In the DM search with transmon qubits, a simple procedure to detect a
DM signal is to observe the excitation of the qubit due to the DM
oscillation.  Taking the initial condition of $\ket{\psi(0)}=\ket{g}$,
the excitation probability from the ground state to the excited state
is given by $|\bra{e}\psi (\tau)\rangle|^2 \simeq \sin^2 \delta \simeq
\delta^2$, where, in the last equality, we have assumed that
$\delta\ll 1$.  If there is no noise nor read-out error, observation
of the excitation of the qubit is a striking signal of the DM coupled
to the qubit.  The process of the DM search proposed in
Ref.\ \cite{Chen:2022quj} is as follows.  Assuming $n_{\rm q}$ qubits
are available, all the qubits are set to the ground state, are left
for evolution for time interval $t=\tau$, and are read out to find the DM
signal; such a cycle of the reset, evolution, and readout processes is
repeated as many times as possible to enhance the sensitivity.  Then,
the search is repeated for different qubit frequencies.

In order to enhance the sensitivity, it is better to have a large number of sensor qubits.  In the above-mentioned procedure, the number of signals scales as $\propto
n_{\rm q}$, assuming that each qubit is individually readout; the probability to observe at least one excited qubit
after each cycle is given by
\begin{align}
  P_{g\rightarrow e}^{\rm (indiv.~readout)}
  \simeq n_{\rm q} \delta^2,
  \label{P(g->e)}
\end{align}
where $n_{\rm q} \delta^2\ll 1$ is assumed.

\vspace{1mm}
\noindent
\section{Quantum circuit}
\label{sec:state}

\begin{figure*}[t]
  \begin{quantikz}
    \lstick[6]{sensors, $\ket{g}^{\otimes n_{\rm q}}$} \slice[style=black]{$t_{\rm i}$} && \targ{}  &  & && \slice[style=black]{$t_{\rm 1}$} & \gate[1][2 cm]{U_{\rm DM}} \slice[style=black]{$t_{\rm 2}$} & && &   &  \targ{ } \slice[style=black]{$t_{\rm f}$} &\meter{} 
    \\
     &\gate{H}& \ctrl{-1} & \targ{}& & &&\gate[1][2 cm]{U_{\rm DM}} & & & & \targ{} & \ctrl{-1}&
    \\
      &\gate{H}& &\ctrl{-1} & \targ{}\vqw{1} & &&\gate[1][2 cm]{U_{\rm DM}} & & &  \targ{}\vqw{1} & \ctrl{-1} &&
    \\
    \wave&&&&&&&&&&&&&
    \\
      &\gate{H}& &&& \ctrl{-1} & \targ{} & \gate[1][2 cm]{U_{\rm DM}}&\targ{}& \ctrl{-1} \ & & & & 
     \\
      &\gate{H}& && && \ctrl{-1} & \gate[1][2 cm]{U_{\rm DM}} &\ctrl{-1} & & & & &
\end{quantikz}
  \caption{Quantum circuit for the DM detection. The gate with $H$ represents the Hadamard gate, while that with ``$\bullet$'' and ``$\oplus$'' connected by the line is the CNOT gate (where ``$\bullet$'' is the control qubit). The $U_{\rm DM}$ represents the evolution with the effect of DM.}
  \label{fig:circuit}
\end{figure*}
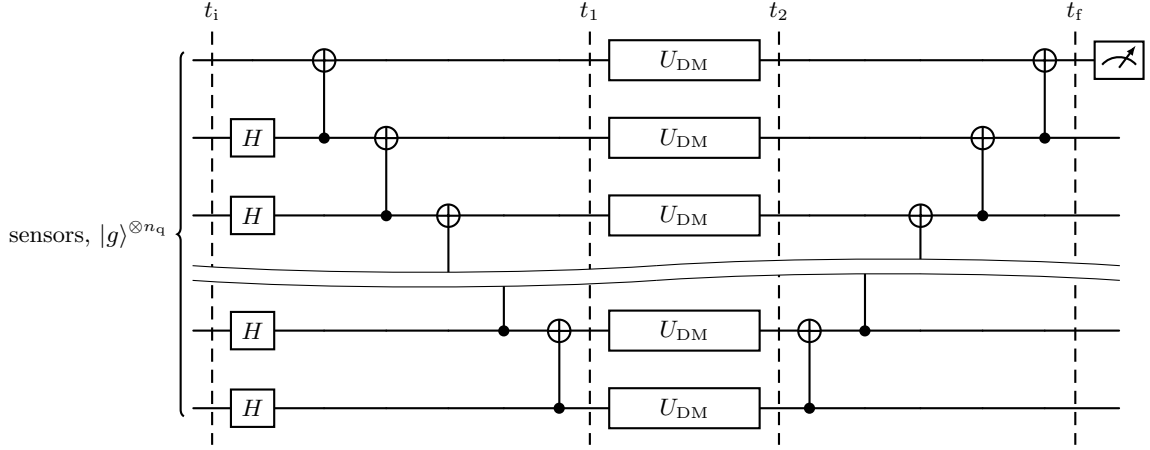

Now, we discuss how the signal rate can be significantly enhanced when $n_{\rm q}\gg 1$.  
We consider a quantum circuit
consisting of $n_{\rm q}$ sensor qubits which interact with DM
oscillation. The state can be expressed by a linear combination of
states of the following form:
\begin{align}
\ket{\Psi}= 
    |\psi_1\rangle \otimes |\psi_2\rangle \otimes 
    \cdots \otimes |\psi_{n_{\rm q}}\rangle,
\end{align}
where $|\psi_i\rangle$ denotes the state of $i$th qubit. 
The frequencies of the sensor qubits are assumed to be
all equal. In our discussion, we consider the case that qubits used
for the DM detection can be initialized, measured, and evolved through
standard gates like the Hadamard gate and CNOT gate. (See Appendix
\ref{appendix:notation} for the gate operations used in our analysis.)

An example of quantum circuits for detecting the DM signal is shown in Fig.\ \ref{fig:circuit}.
This is a quantum circuit for quantum-enhanced parameter estimation\,\cite{QuantumEnhancedMeasurement, NaturePhoton5-222, PRL115-170801}. Our circuit consists of only one-dimensional nearest neighbor interaction between qubits with $\mathcal{O}(n_{\rm q})$ gates. 
We assume that $t_1-t_{\rm i} \sim t_{\rm f}-t_2 \ll t_2-t_1$, so that the effect of DM is mainly in time interval $t_1 \leq t \leq t_2$. We also assume that the coherence time of the qubits is long enough, so that the coherence time of the system, $\tau$, is determined by the coherence of the DM and does not scale with $n_\text{q}^{-1}$. We expect that the coherence time of the qubit system longer than that of DM is achievable in future quantum computer platforms with sizable $n_{\rm q}$. The entangled qubit system is usually more fragile than the individual nonentangled ones and the coherence time of the entangled state may be $\sim\tau_{\rm q}/n_{\rm q}$, where $\tau_{\rm q}$ is the coherence time of a single qubit \cite{PRL79-3865}.  Even in such a case, the following discussion holds as far as $n_{\rm q}\lesssim \tau_{\rm q}/\tau_{\rm DM}$ (with $\tau_{\rm DM}$ being the coherence time of the DM).

In order to understand the enhancement mechanism of the signal, it is instructive to consider the case that $\alpha=0$. 
For $\alpha=0$, the eigenstates of $U_{\rm DM}$ are $\ket{+}$ and $\ket{-}$, satisfying $U_{\rm DM}\ket{\pm}=e^{\pm i\delta}\ket{\pm}$, where
\begin{align}
  \ket{\pm} \equiv \frac{1}{\sqrt{2}} ( \ket{g} \pm \ket{e} ).
\end{align}
Thus, considering the states with $n_{\rm q}$ qubits, $\ket{\pm}^{\otimes n_{\rm q}}$, they evolve as $\ket{\pm}^{\otimes n_{\rm q}} \rightarrow U_{\rm DM}^{\otimes n_{\rm q}}\ket{\pm}^{\otimes n_{\rm q}}=e^{\pm i n_{\rm q} \delta} \ket{\pm}^{\otimes n_{\rm q}}$; the phases from $n_{\rm q}$ qubits coherently add up. 
Our quantum circuit measures this phase as the relative phase between $\ket{+}^{\otimes n_{\rm q}}$ and $\ket{-}^{\otimes n_{\rm q}}$ by using the superposition of these states.

With the circuit, the state evolves as follows.  First, all the qubits are prepared in the ground state at $t=t_{\rm i}$. At $t=t_1$,  the state of sensor qubits is given by
\begin{align}
  \ket{ \Psi (t_1) } = 
  \frac{1}{\sqrt{2}}
  \left( 
  \ket{+}^{\otimes n_{\rm q}} +
  \ket{-}^{\otimes n_{\rm q}}
  \right).
  \label{psi_init}
\end{align}
With the effect of the DM, the state at $t=t_2$ becomes
\begin{align}
  \ket{\Psi (t_2)} = 
  &\frac{1}{\sqrt{2}}
  \left( 
  e^{in_{\rm q}\delta} \ket{+}^{\otimes n_{\rm q}} +
  e^{-in_{\rm q}\delta} \ket{-}^{\otimes n_{\rm q}}
  \right).
 \label{eq:psi_t2}
\end{align}
The quantum operation from $t=t_2$ to $t_{\rm f}$ brings the phase information to the first qubit:
\begin{align}
  \ket{\Psi (t_{\rm f})} &= \frac{1}{\sqrt{2}}
  \left( e^{in_{\rm q} \delta}\ket{+}+ e^{-in_{\rm q} \delta}\ket{-} \right)\otimes \ket{+}^{\otimes (n_{\rm q}-1)}
  \nonumber \\
  &= \left[ \cos(n_{\rm q} \delta) \ket{g}+ i\sin(n_{\rm q} \delta) \ket{e} \right] \otimes \ket{+}^{\otimes (n_{\rm q}-1)}.
 \label{eq:psi_tf}
\end{align}
The probability to observe the excitation of the first qubit is
\begin{equation}
  P^{(\alpha=0)}_{g\rightarrow e}= \sin^2 (n_{\rm q} \delta) \simeq n_{\rm q}^2 \delta^2,
  \label{P(g->e)alphaknown}
\end{equation}
where, in the last equality, we have used $n_{\rm q} \delta \ll 1$. Notably, the probability is proportional to $n_{\rm q}^2$, indicating a possible enhancement of the signal using the quantum properties of the qubits.

We can use our circuit even in actual situations where $\alpha$ is unknown. Concentrating on the case that $\delta\ll 1$, we may expand the evolution operator for $n_{\rm q}$ qubits as
\begin{align}
  U_{\rm DM}^{\otimes n_{\rm q}} \simeq {\bf 1} + 
  i \delta \sum_{i=1}^{n_{\rm q}}
  \left( X_i \cos\alpha + Y_i \sin\alpha \right)
  + \mathcal{O}(\delta^2),
\end{align}
where the summation is over the operators acting on all the qubits. For any $i$, the following relation holds:
\begin{align}
    X_i \ket{\pm}^{\otimes n_{\rm q}} =
    \pm \ket{\pm}^{\otimes n_{\rm q}},
  \label{Xi|pm>}
\end{align}
with which we obtain
\begin{align}
  \ket{\Psi (t_2)} \simeq &\,
  \frac{1}{\sqrt{2}}
  \left( 1 + i n_{\rm q} \delta \cos\alpha \right)
  \ket{+}^{\otimes n_{\rm q}}
  \nonumber \\ &\, 
  + \frac{1}{\sqrt{2}}
  \left( 1 - i n_{\rm q} \delta \cos\alpha \right)
  \ket{-}^{\otimes n_{\rm q}} + \cdots,
  \label{Psi(t2):alpha}
\end{align}
where terms irrelevant to our discussion are neglected. The relevant part of Eq.\ \eqref{Psi(t2):alpha} is obtained from Eq.\ \eqref{eq:psi_t2} with replacing $\delta\rightarrow\delta\cos\alpha$; the excitation probability of the first qubit is found to be
\begin{equation}
  P_{g\rightarrow e}^{\rm (\alpha:unknown)} \simeq
  n_{\rm q}^2 \delta^2 \cos^2 \alpha + \mathcal{O}(n_\text{q}\delta^2).
  \label{eq:P01}
\end{equation}
In the proposed DM detection experiment~\cite{Chen:2022quj},
the cycle of the reset, evolution, and readout processes is repeated many times.
At each cycle, $\alpha$, the phase of the DM oscillation, is unknown and can be regarded as random. 
We can thus take the average over $\alpha$; for $n_{\rm q} \gg 1$, the excitation probability is
\begin{align}
  P_{g\rightarrow e}
  \simeq \frac{1}{2} n_{\rm q}^2 \delta^2. \label{eq:sigrate}
\end{align}
One can see that $P_{g\rightarrow e}$ vanishes for the case without the DM (i.e., $\delta=0$). Thus, the detection of the excitation of the first qubit can be regarded as a signal of the DM.  More importantly, the signal rate is proportional to $n_{\rm q}^2$ as opposed to $n_{\rm q}$ thanks to the coherent phase evolution of the entangled state.

Lastly, we make a few comments about our quantum circuit. First, our circuit requires $\mathcal{O}(n_\text{q})$ steps to prepare the initial state, Eq.\,\eqref{psi_init}. However, with additional qubits, we may prepare the initial state with $\mathcal{O}(1)$ steps. The detail is discussed in Appendix \ref{appendix:quantumnoise} and \ref{appendix:error_correc}. Second, the quantum circuit between $t_2\leq t\leq t_{\rm f}$ can be regarded as the parity measurement of $\ket{\Psi(t_2)}$, where the parity $P\equiv\prod_i Z_i$ is the product of the $Z$ operator of each qubit.  (Notice that, under the parity operation, $\ket{+}^{\otimes n_{\rm q}}\overset{P}{\leftrightarrow}\ket{-}^{\otimes n_{\rm q}}$.)

\vspace{1mm}
\noindent
\section{Effect of noises}
Next, we evaluate the tolerance to noises in our quantum circuit.  
The quantum noises can be associated with either the gate operations or the evolution and may result in dark counts. Thermal noise may also mimic the signal. When the noise rate is low enough, the number of dark counts is expected to scale linearly with the number of qubits and gate operations in the circuit, i.e., $\mathcal{O}(n_{\rm q})$, where the impact of the dark counts on the signal-to-noise ratio is subdued with a sufficiently large $n_{\rm q}$.
The assumption is not always trivial when a large number of qubits are involved in an entangled state since it is generally more susceptible to decoherence through de-phasing errors.
In the following discussion, however, we find that our sensor qubits are robust against de-phasing errors,
while other errors such as de-excitation or bit-flip remain inevitable.

We first consider the case without $U_\text{DM}$ to study what type of noises mimic the signals, yielding dark counts.
Noises during $t_1 < t < t_2$ are analyzed with a given state Eq.\,\eqref{psi_init} at $t = t_1$.
As discussed in Appendix~\ref{appendix:quantumnoise}, noises are described as a set of Kraus operators $\{E_k\}$, where $k$ indicates the type of operators; an initial pure state $\ket\psi$ is mapped onto a classical mixture of states $\{E_k|\psi\rangle\}$,
where we ignore normalization factors in this section for simplicity.
Assuming the noise influences each qubit independently, $E_k$ is an operator on a single qubit space. In the following, we focus on noises described by a single Kraus operator $E$.

First, we consider $E = X_i$, which corresponds to a bit-flip noise. With this error, the state at $t=t_2$ becomes
\begin{align}
    X_i \ket{\Psi(t_1)} = \frac{1}{\sqrt{2}}
  \left( 
  \ket{+}^{\otimes n_{\rm q}} -
  \ket{-}^{\otimes n_{\rm q}}
  \right),
\end{align}
which is indistinguishable from the real excitation signals by the DM.
The state at $t=t_{\rm f}$ is therefore
\begin{align}
    \ket{\Psi(t_{\rm f})}=  \ket{e} \otimes \ket{+}^{\otimes (n_{\rm q}-1)},
\end{align}
faking the signals with the first qubit being excited. 
Note that the quantum error correction algorithms do not ameliorate this problem since the error and the signal are reduced to the same state.
Similarly, errors involving $X_i$ or $Y_i$, such as de-excitation channel, $E = X_i + iY_i$, mimic the DM signal and cannot be corrected.
The result is not surprising because these errors matter even when the entanglement of qubits is not used\,\cite{Chen:2022quj}.

Next, we consider $E = Z_i$. This corresponds to the de-phasing noise. The state at $t=t_2$ is
\begin{align}
  Z_i\ket{\Psi(t_1)} =  
  \frac{1}{\sqrt{2}} & 
  \left[  
    \ket{+}^{\otimes (i-1)} \otimes \ket{-} \otimes \ket{+}^{\otimes (n_{\rm q}-i)} 
    \right. 
    \nonumber \\ &
    \left. +  \ket{-}^{\otimes (i-1)} \otimes \ket{+} \otimes \ket{-}^{\otimes (n_{\rm q}-i)}  \right],
\end{align}
where $i$-th qubit is transformed as $\ket{\pm}\rightarrow\ket{\mp}$, resulting in:
\begin{equation}
    \ket{\Psi (t_f)} 
    =\ket{g} \otimes \left[  \ket{+}^{\otimes (i-2)} \otimes \ket{-} \otimes \ket{+}^{\otimes (n_{\rm q}-i)}\right].
\end{equation}
One finds the first qubit remains in the ground state, indicating that the error $Z_i$ does not induce the dark counts.

The discussion almost applies in the same way when $U_{\rm DM}$ is taken into account. We elaborate on the de-phasing channel which is nontrivial among noises given above. 
For simplicity, let us assume $\alpha=0$ and that only one shot of error $E=Z_i$ occurs at $t = t' \in [t_1, t_2]$.
The state at $t=t_2$ is:
\begin{align}
  \ket{\Psi(t_2)} =  
  \frac{1}{\sqrt{2}} & 
  \left[  
    e^{in_{\rm q}\delta'}\ket{+}^{\otimes (i-1)} \otimes \ket{-} \otimes \ket{+}^{\otimes (n_{\rm q}-i)} 
    \right. 
    \nonumber \\ &
    \left. +  e^{-in_{\rm q}\delta'}\ket{-}^{\otimes (i-1)} \otimes \ket{+} \otimes \ket{-}^{\otimes (n_{\rm q}-i)}  \right],
\end{align}
where $n_{\rm q}\delta'\equiv n_\text{q}\eta\tau' + (n_\text{q} - 2)\eta(\tau - \tau')$ and $\tau' = t' - t_1$. 
While the signal rate is penalized according to the number of qubits affected by the de-phasing noise, it is remarkable that it does not break the multi-bit coherence, and that the $\mathcal{O}(n_\text{q}^2)$ excitation probability is still maintained. 
This de-phasing error insensitivity can be understood by seeing this quantum circuit as an error correction code, the details of which we elaborated in Appendix~\ref{appendix:error_correc}. There, we also show that we can even identify and correct de-phasing errors, which is helpful when a large number of qubits are affected by the error during $t_1 \leq t \leq t_2$.

\vspace{1mm}
\noindent
\section{Conclusions and discussion}
We present a signal enhancement method for the DM detection experiment using qubits proposed in our earlier work~\cite{Chen:2022quj}. Employing a tailored kernel of a quantum circuit, the signal rate can be made proportional to $n_{\rm q}^2$ as opposed to $n_{\rm q}$, reflecting the nature of quantum interference. We have provided a specific example of the circuit (see Fig.\ \ref{fig:circuit}) that realizes such an enhancement. 
The circuit is feasible with the architecture of a one-dimensional chain of qubits, and found to be fault tolerant against the de-phasing noise.

One of the promising candidate platforms to host such an experiment is the superconducting quantum computer with an array of transmon qubits, as discussed in Ref.\ \cite{Chen:2022quj}. While it allows probing the unexplored parameter region of hidden photon DM even without the proposed quantum enhancement, a significant improvement is expected with the enhancement. With the current best superconducting qubit technology, a longitudinal coherence time of $\tau_{\rm q}\sim 1\ {\rm ms}$ and a two-qubit gate error rate of $\sim 0.5\ \%$ are achieved. Given that the coherence time of the entangled $n_{\rm q}$-qubit system scales as $\sim \tau_{\rm q}/n_{\rm q}$ and that the DM coherence time is $\tau_{\rm DM}\sim 0.1\ {\rm ms}$ for the case of our interest, we may use $n_{\rm q}\sim 10$ entangled qubits with satisfying $\tau_{\rm DM}\lesssim\tau_{\rm q}/n_{\rm q}$. The two-qubit gate error rate defines the level of dark count. Assuming $n_{\rm q}=10$, a two-qubit gate error rate of $0.5\ \%$, and other circuit parameters adopted in Ref.\ \cite{Chen:2022quj}, $\epsilon$ down to $O(10^{-13})$ can be probed with the proposed entanglement scheme. To go further, improvements are required in the coherence time of the qubits and also in gate fidelities (particularly for two-qubit gates like CNOT). The requirements above are shared with the future fault-tolerant quantum computers (FTQCs), and hence we expect that the technical challenges will be significantly lowered on the way towards FTQCs being realized. Assuming that $\tau_{\rm q}=10\ {\rm ms}$ and a two-qubit gate error rate of $0.1\ \%$ can be realized, 
the sensitivity reaches $\epsilon$ down to $O(10^{-14})$ using $n_{\rm q} \sim 100$. These specifications can be further improved, and the gain is open-ended.

Towards the practical implementation, one challenge may be that the qubit frequencies have to be aligned to the DM mass within their intrinsic widths, and be able to scan over multiple-GHz range.
Since this is a highly anomalous configuration for quantum computers to operate, it is not likely possible to experiment in a parasitic manner during normal operation.
However, with a frequency tunable computer (e.g., Google Sycamore~\cite{arute2019quantum}) the frequency alignment and the scan are technically feasible.

The other consideration involves the package coefficient $\kappa$ in Eq.\ \eqref{eq:eta}.
In conventional superconducting quantum computers, the quantum processor chips are typically covered with a hermetic metal shield to minimize qubits' spontaneous radiation as well as the external radiation noises, which can greatly suppress $\kappa$ below 1.

Quantum computers using other modalities, especially trapped ions and Rydberg atoms, can potentially circumvent these issues.
Although their coupling with photons is generally weaker than that of superconducting qubits, they are advantageous in that the frequency alignment is ensured by nature, and that they are free from the shielding issue.
The better frequency alignment also allows us to entangle a larger number of qubits using the proposed quantum enhancement, which can compensate for the disadvantage in the signal rate due to the weaker coupling.
These factors enable the potential of the DM detection experiment to be performed on the same hardware as the current or future quantum computer machines.

\noindent
\section*{Acknowledgments}
S. C. was supported by JSPS KAKENHI Grants No.\ 23K13093. T. I. was supported by JSPS KAKENHI Grants No.\ 23H04864 and JST PRESTO Grant No.\ JPMJPR2253, Japan. T. M. was supported by JSPS KAKENHI Grants No.\ 22H01215 and No.\ 23K22486. T. N. was supported by JSPS KAKENHI Grants No. 23K17688, No. 23H01182, and JST PRESTO Grant No.\ JPMJPR23F7, Japan. T. S. was supported by the JSPS fellowship Grant No.\ 23KJ0678. 

\appendix

\section{Notation} \label{appendix:notation}
\setcounter{equation}{0}
The following are definitions adopted for gates that appeared in this Letter.
\begin{itemize}
\item Hadamard gate $H$:
  \begin{align}
    \begin{pmatrix}
      | g \rangle \\ |e \rangle
    \end{pmatrix}
    \xrightarrow{H}
    \frac{1}{\sqrt{2}}
    \begin{pmatrix}
      1 & 1 \\ 1 & -1
    \end{pmatrix}
    \begin{pmatrix}
      | g \rangle \\ |e \rangle
    \end{pmatrix}.
  \end{align}
\item $X$, $Y$, and $Z$ gates:
  \begin{align}
    \begin{pmatrix}
      | g \rangle \\ |e \rangle
    \end{pmatrix}
    \xrightarrow{X, Y, Z}
    \sigma_{X, Y, Z}        
    \begin{pmatrix}
      | g \rangle \\ |e \rangle
    \end{pmatrix},
  \end{align}
    where $\sigma_X$, $\sigma_Y$, and $\sigma_Z$ are Pauli's $\sigma$ matrices.  $X$, $Y$, and $Z$ are also used for operators acting on qubits.  

\item Controlled NOT gate (CNOT):
  \begin{align}
    \begin{pmatrix}
      |g \rangle \otimes | g \rangle \\
      |g \rangle \otimes | e \rangle \\
      |e \rangle \otimes | g \rangle \\
      |e \rangle \otimes | e \rangle
    \end{pmatrix}
    \xrightarrow{{\rm CNOT}}
    \begin{pmatrix}
      1 & 0 & 0 & 0 \\
      0 & 1 & 0 & 0 \\
      0 & 0 & 0 & 1 \\
      0 & 0 & 1 & 0 
    \end{pmatrix}
    \begin{pmatrix}
      |g \rangle \otimes | g \rangle \\
      |g \rangle \otimes | e \rangle \\
      |e \rangle \otimes | g \rangle \\
      |e \rangle \otimes | e \rangle
    \end{pmatrix},
  \end{align}
  where the first qubit is the control qubit and the second is the target qubit.

\end{itemize}
For more details about quantum gates and quantum circuits, see, for
example, Ref. \cite{Nielsen:2012yss}.

\section{Quantum operation and quantum noise} \label{appendix:quantumnoise}
\setcounter{equation}{0}

We review the basics of quantum operation and how quantum noises are taken into account as quantum operations. Good reviews are found in Refs.\,\cite{PreskillLecture,Nielsen:2012yss}.

A state in quantum mechanics can be described by a density matrix operator, $\rho$. Any evolution of the state is described by an operation called the quantum operation, $\mathcal{E}$; a quantum state $\rho$ is mapped into $\mathcal{E}(\rho)$. Elementary examples of $\mathcal{E}$ include unitary evolution and measurements.
As the mapped operator $\mathcal{E}(\rho)$ is also a density matrix, $\mathcal{E}$ must satisfy three conditions. First, it must conserve the trace.
Second, the map of a classical mixture of density matrices $\rho_i$ with probability $p_i$ must be equal to a classical superposition of the mapped density matrices $\mathcal{E}(\rho_i)$ with probability $p_i$. Lastly, the eigenvalues of the mapped operator $\mathcal{E}(\rho)$ must be positive; in addition, if we introduce an additional system $R$ and perform a composed quantum operation $\mathcal{I}_R \otimes \mathcal{E}$, where $\mathcal{I}_R$ is the identity operation in $R$, on a density matrix $\rho$ of the composed system, the eigenvalues of $(\mathcal{I}_R \otimes \mathcal{E})(\rho)$ must also be positive.

With these conditions, it is known that any quantum operation can be written as the operator sum representation; any $\mathcal{E}$ can be written as
\begin{align}
    \mathcal{E}(\rho) = \sum_k E_k \rho E_k^\dagger,
\end{align}
where operators $\{E_k\}$ are called the Kraus operators. The set of Kraus operators conserves the trace of density matrices;
\begin{align}
    \sum_k E_k^\dagger E_k = \mathbf{1}.
\end{align}
Unitary evolution and measurements have the operator sum representation.
Note that the operator sum representation is not unique, as $F_i = U_{ij} E_j$, where $U$ is a unitary matrix, describes the same map on density matrices.

Let us get an insight into the physical meaning of the quantum operation and Kraus operators. Without loss of generality, we may start with a pure state, $\rho =|\psi\rangle\langle\psi|$ for some state $\ket{\psi}$, as mixed states are a classical mixture of pure states. After a quantum operation $\mathcal E$, the state is mapped to $\sum E_k |\psi\rangle\langle\psi| E_k^\dagger$. The state is understood as a classical mixture of states $\left\{\frac{E_k \ket{\psi}}{|E_k \ket{\psi}|}\right\}$ with probabilities $\{|E_k |\psi\rangle|^2\}$. Thus, we can interpret that a quantum operation $\mathcal{E}$ maps a state $\rho$ into a state $\rho_k \equiv E_k \rho E_k^\dagger / \text{Tr}(E_k \rho E_k^\dagger)$ with a probability $\text{Tr}(E_k \rho E_k^\dagger)$.

As any map on a density matrix can be written using a quantum operation, quantum noises on a qubit are also written as quantum operations. In general, any Kraus operator on one qubit can be expanded in $I, X, Y$, and $Z$, where $I$ is the identity operator:
\begin{align}
  E_k = c^k_I I + c^k_X X + c^k_Y Y + c^k_Z Z.
  \label{Kraus}
\end{align}
Among various errors on one qubit, the de-excitation and de-phasing channels are of particular importance; the sets of Kraus operators of the de-excitation and de-phasing are
\begin{align}
    E_0^A &= \begin{pmatrix}
        1 & 0 \\
        0 & \sqrt{1 - p_A}
    \end{pmatrix} 
    = \frac{1 + \sqrt{1 - p_A}}{2} I + \frac{1 - \sqrt{1 - p_A}}{2} Z, \\
    E_1^A &= \begin{pmatrix}
        0 & \sqrt{p_A} \\
        0 & 0
    \end{pmatrix}
    = \frac{\sqrt{p_A}}{2} X + i \frac{\sqrt{p_A}}{2} Y,
\end{align}
and 
\begin{align}
    E_0^{P'} &= \begin{pmatrix}
        1 & 0 \\
        0 & \sqrt{1 - p_{P'}}
    \end{pmatrix}, \\
    E_1^{P'} &= \begin{pmatrix}
        0 & 0 \\
        0 & \sqrt{p_{P'}}
    \end{pmatrix},
\end{align}
respectively, where $p_A$ and $p_{P'}$ governs the damping probability of the amplitude and phase, respectively, and satisfy $0 \le p_A, p_{P'} \le 1$. By the unitary transformation described above, the Kraus operators of the phase damping can be rewritten as
\begin{align}
    E_0^P &= \sqrt{1 - p_P} I, \\
    E_1^P &= \sqrt{p_P} Z,
\end{align}
with the relation $p_P = (1-\sqrt{1-p_{P'}})/2$.
Therefore, the channel is also called the phase flip channel.

\section{Error correction} \label{appendix:error_correc}
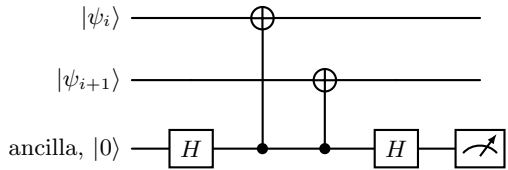
\begin{figure}[t]
  \centering
  \begin{quantikz}
        \lstick{\ket{\psi_i}}      &  & \targ{} &     & &\\
        \lstick{\ket{\psi_{i+ 1}}} & &          & \targ{} &  &\\
      \lstick{ancilla, \ket{0}}    & \gate{H} &  \ctrl{-2} & \ctrl{-1}  & \gate{H}&\meter{}
  \end{quantikz}
  \caption{Quantum circuit for the syndrome $X_iX_{i+1}$ measurement.}
  \label{fig:synd_circuit}
\end{figure}

\setcounter{equation}{0}
We describe that our circuit can be regarded as an error correction code, illustrating why the enhanced signal is protected by de-phasing errors.
First, let us suppose that $\alpha=0$.
The states evolve in the Hilbert subspace spanned by $\{\ket{+}^{\otimes n_{\rm q}}, \ket{-}^{\otimes n_{\rm q}}\}$, and this is invariant under the group of operations of $S = \{X_1X_2, X_2X_3, \dots, X_{n_\text{q} - 1}X_{n_\text{q}}\}$. Such a subspace corresponds to a stabilizer code with the generating set $S$ and can also be regarded as one column of surface code\,\cite{Bravyi:1998sy,Dennis:2001nw,Fowler_2012}. This implies that we can even identify and correct de-phasing errors without breaking the coherence, which is helpful when a large number of qubits are affected by the error during $t_1 \leq t \leq t_2$.
To identify the errors, we repeatedly measure the eigenvalues of ``error syndrome operators'' $\{X_iX_{i+1}\}$ during $t_1 \leq t \leq t_2$, and examine the change of outcomes between each cycle of measurements. 
The measurement of $X_iX_{i+1}$ can be performed with an ancilla qubit using, e.g., the circuit shown in Fig.\ \ref{fig:synd_circuit}. All the syndrome operators return $+1$ without the de-phasing error, while two syndromes, $X_{i - 1}X_i$ and $X_i X_{i + 1}$, return $-1$ in case of a single $E = Z_i$ being in action. Subsequently, we recover the state by applying $Z$ gates accordingly. (We assume that the gate and measurement errors are negligible.)

In reality, due to the effect of the DM with unknown phase $\alpha$, the state is not completely confined within the subspace  $\{\ket{+}^{\otimes n_{\rm q}}, \ket{-}^{\otimes n_{\rm q}}\}$ even without the $Z_i$ error, i.e., the state is not an eigenstate of the syndrome operators.
However, the coherence within the subspace $\{\ket{+}^{\otimes n_{\rm q}}, \ket{-}^{\otimes n_{\rm q}}\}$ is maintained and Eq.\,\eqref{eq:P01} still holds even with the error correction.

Finally, we comment that, with syndrome measurement and subsequent feedback, we can prepare the state at $t=t_1$, Eq.\,\eqref{psi_init}, in $\mathcal{O}(1)$ steps of gate operations. This can be seen as follows. 
For simplicity, let us assume that $n_{\rm q}$ is odd. 
Suppose that initially all qubits are prepared in the ground state.  Using $\ket{g}=\frac{1}{\sqrt{2}}(\ket{+}+\ket{-})$, as well as $Z\ket{+}=\ket{-}$ and $Z^2 = 1$, we can rewrite the state as:
\begin{widetext}
\begin{align}
\ket{g}^{\otimes n_{\rm q}}
 =&\left(\frac{1}{\sqrt{2}}\right)^{n_{\rm q}} \sum_{\{q_{i}=0,1\}} Z^{q_1}_1Z^{q_2}_2...Z^{q_{n_{\rm q}}}_{n_{\rm q} }\ket{+}^{\otimes n_{\rm q}} \nonumber\\
 =&\left(\frac{1}{\sqrt{2}}\right)^{(n_{\rm q}-1)} \left[ \sum_{\{q_{i}=0,1\}}^{\sum_i q_i<n_{\rm q}/2} Z^{q_1}_1Z^{q_2}_2...Z^{q_{n_{\rm q}}}_{n_{\rm q} } \left( \frac{\ket{+}^{\otimes n_{\rm q}} + Z_1 Z_2... Z_{ n_{\rm q}} \ket{+}^{\otimes n_{\rm q}}}{\sqrt{2}} \right) \right] \nonumber \\
=&
\left(\frac{1}{\sqrt{2}}\right)^{(n_{\rm q}-1)} \left[ \sum_{\{q_{i}=0,1\}}^{\sum_i q_i<n_{\rm q}/2} Z^{q_1}_1Z^{q_2}_2...Z^{q_{n_{\rm q}}}_{n_{\rm q} } \left( \frac{\ket{+}^{\otimes n_{\rm q}} + \ket{-}^{\otimes n_{\rm q}}}{\sqrt{2}} \right) \right].
\label{g^n}
\end{align}
\end{widetext}
Notice that all syndrome operators return $+1$ for $\frac{1}{\sqrt{2}}( \ket{+}^{\otimes n_{\rm q}} + \ket{-}^{\otimes n_{\rm q}} )$ and that $X_i X_{i+1}$ anticommutes with $Z_i$ and $Z_{i+1}$.  Thus, all states in the linear combination in the last line of Eq.\ \eqref{g^n} are distinguished by the eigenvalues of the syndrome operators. This implies that, with the syndrome measurement, the state collapses into one of the states in the following form:
\begin{equation}
Z^{q_1}_1Z^{q_2}_2...Z^{q_{n_{\rm q}}}_{n_{\rm q} } \left( \frac{\ket{+}^{\otimes n_{\rm q}} + \ket{-}^{\otimes n_{\rm q}}}{\sqrt{2}} \right),
\end{equation}
of which the value of all $q_i$ is known by the measurement outcome.
By applying $Z$ gates accordingly we can prepare the desired state at $t=t_1$, Eq.\,\eqref{psi_init}.
The argument applies similarly when $n_{\rm q}$ is even.

\bibliographystyle{apsrev4-1}
\bibliography{ref.bib}


\end{document}